# THE ELUSIVITY OF NATURE AND THE MIND-MATTER PROBLEM

Brian D. Josephson
Cavendish Laboratory, Madingley Road, Cambridge CB3 0HE, England.

ABSTRACT

This paper examines the processes involved in attempting to capture the subtlest aspects of nature by the scientific method and argues on this basis that nature is fundamentally elusive and may resist grasping by the methods of science. If we wish to come to terms with this resistance, then a shift in the direction of taking direct experience into account may be necessary for science's future complete development.

---------------------------------------

What is matter and what is mind? René Descartes regarded these entities as being of totally different orders and completely separate from each other, although nowadays the conventional view is to consider mind function as reducible to the behaviour of matter. In this paper I want to propose a new theme in the mind-matter problem, with an approach that considers what I shall call the elusivity or elusive quality of nature. The fundamental idea is that nature at large is less amenable to being pinned down to a precise description than the successes of science to date may seem to imply, particularly if we consider science as applying to some 'objective reality' that is independent of the scientist. Some phenomena are more elusive to the methods of ordinary science than others, as will be explained in terms of a series of examples in what follows. Mind may be especially elusive and it will be suggested in the course of this paper that as far as the phenomena of mind go, there is no good reason to suppose that a proper description in terms of the usual methods of science can be obtained at all. To deal with this situation it may prove necessary to use subtler methodologies, such as entering into the subtleties of conscious experience directly, as is the way of the mystic or of the practitioner of the arts.

To illustrate the concept of elusivity, let us focus on the problem of knowing precisely where a given object is located. The least elusive entities are classical macroscopic objects. The position of such an object relative to a given reference frame can be typically measured by means of a ruler, or by means of the techniques of the surveyor, depending on on the length scales involved (or by more advanced methods such as those involving lasers if high accuracy is desired). If, however, the object is moving then the question of its position becomes problematic, as in the case of the Zeno paradox. If the velocity is very low then the concept of the current position has meaning. However, if the motion is rapid the concept of 'now' is not well enough defined to make the position

'now' meaningful. In such a case the concept of the time (coupled with the physical apparatus of the clock that gives a measure to this variable) becomes necessary in order to be able to determine and define the position of the object at particular instant of time which is of interest. Defining this elusive quantity (the position) requires the use of additional apparatus such as a high-speed shutter or light source to 'freeze' the position and make it definite.

In the microscopic domain, we never determine the position of a high-speed fundamental particle at a given moment of absolute time at all, but only measure its trajectory. In the quantum domain, elusiveness becomes a fundamental feature of nature since the uncertainty principle sets definite limits to our ability (even in principle) to determine the position of a particle over a period of time of finite duration. Nature in this regime behaves rather like an insect that reponds to our attempts to swat it, utilising a determination of just where it will be at the appropriate future instant, by changing its position. This ensures that our prediction will be incorrect, with the result that the insect survives our attempts to dispose of it.

Arguments can be given (Josephson 1988) that in the real world (i.e. the world outside the domain of the scientific laboratory) nature may be still more elusive than this. Detailed analysis and discussion of arguments due originally to Niels Bohr would appear to show that in general any attempt to discover the essence of a naturally occurring system in such detail as to be able to describe it within a quantum mechanical framework (viz. by assigning to it a wave function or a density matrix) will lead to such an alteration of the characteristics of the system under study that it will no longer be possible to regard it as being applicable to the original system. Again, it may be noted that the phenomenon of deterministic chaos (Gleick 1988) points equally to the existence of unavoidable limitations to the attempt to pin down nature and say exactly what it is.

Thus some aspects of nature are more amenable to the way of present-day science than are others. We can start to see mind, in its subtler aspects especially, as something which cannot necessarily be handled in terms of the kind of methods and descriptions that we are used to using in the world of science. Even if it is in fact legitimate to equate the workings of the mind with the behaviour of the brain, the question of what exactly the brain is will present us with grave difficulties, even in principle, if we ask questions about it at a sufficiently subtle level.

But these traditional methods of science, at least for the hard sciences, take little or no account of conscious experience, which is regarded as an epiphenomenon that science does not have to take account of at all. Yet our own experiences are in general less elusive to us than are the subtler phenomena of modern science. It may thus be argued that if we are to understand the mind fully then science needs to take fuller account of the phenomena of conscious experience. In this respect, conscious experience is usually considered to be a subjective affair and thus unsuitable for meeting the requirements of science. I believe this to be a mistaken view of the situation. Artists and mystics have, in some cases, described their conscious experiences in great detail, and I see no reasons (other than custom and ideology) why these should not be taken into account in assembling our world-view. A closely related topic is the question of meaning, which David Bohm has suggested (Pylkkäanen 1989) constitutes a fundamental aspect of reality. Again, in the arts we find meaning to have dimensions that the methods of reasoning employed in science have yet to discover (Langer 1957).

I will conclude by suggesting that if we wish to come to terms with the resistance that the natural world appears to have to being grasped by the methods of science, then a shift in the direction of taking into account direct experience may be necessary for the future development of science, although this will necessarily entail a drastic shift in regard to what it is regarded as being appropriate areas of exploration for the scientist.

I am grateful to Prof. S.M. Rosen for discussions of experiential knowledge.